\pdfoutput=1
\documentclass[conference,a4paper]{IEEEtran}

\usepackage{xcolor}
\usepackage{balance}
\usepackage{cite}
\ifCLASSINFOpdf
  \usepackage[pdftex]{graphicx}
\else
  \usepackage[dvips]{graphicx}
\fi
\graphicspath{{figs/}}
\usepackage[caption=false,font=footnotesize]{subfig}
\usepackage{amsmath,amssymb}
\usepackage{booktabs}
\usepackage{siunitx}
\makeatletter
\newcommand{\linebreakand}{%
  \end{@IEEEauthorhalign}
  \hfill\mbox{}\par
  \mbox{}\hfill\begin{@IEEEauthorhalign}}
\makeatother

\begin{document}

\title{Phase-Matched Design of Pinching-Antenna Elements on a 60-GHz Dielectric Rod Waveguide}

\author{
\IEEEauthorblockN{Stylianos D. Asimonis}
\IEEEauthorblockA{\textit{Dept. of Electrical \& Computer Eng.}\\
\textit{University of Patras}\\
26504 Patras, Greece\\
s.asimonis@upatras.gr}
\and
\IEEEauthorblockN{Dimitrios Tyrovolas}
\IEEEauthorblockA{\textit{Dept. of Informatics \& Telecom.}\\
\textit{University of Peloponnese}\\
22131 Tripoli, Greece\\
d.tyrovolas@uop.gr}
\and
\IEEEauthorblockN{Spyridon Daskalakis}
\IEEEauthorblockA{\textit{Inst. of Sensors, Signals and Systems} \\
	\textit{Heriot-Watt University}\\
	Edinburgh EH14 4AS, U.K. \\
	s.daskalakis@hw.ac.uk}
\linebreakand
\IEEEauthorblockN{Nikos C. Sagias}
\IEEEauthorblockA{\textit{Dept. of Informatics \& Telecom.}\\
\textit{University of Peloponnese}\\
22131 Tripoli, Greece\\
nsagias@uop.gr}
\and
\IEEEauthorblockN{Donal P. Lynch}
\IEEEauthorblockA{\textit{Steatite Antennas}\\
Leominster HR6 0QF, UK\\
dlynch27@qub.ac.uk}
\and
\IEEEauthorblockN{George K. Karagiannidis}
\IEEEauthorblockA{\textit{Dept. of Electrical \& Computer Eng.}\\
\textit{Aristotle University of Thessaloniki}\\
54124 Thessaloniki, Greece\\
geokarag@auth.gr}
}

\maketitle

\begin{abstract}
The radiation of pinching antennas, dielectric elements placed on a dielectric waveguide, depends on the
element shape. We explain it with a phase-matching, or prism-coupler, picture: the guided rod mode
couples into the denser antenna plate at an angle set by the plate-mode effective index, about
31 degrees at 60 GHz, and radiates through the face it reaches first. This yields a simple sizing rule linking
element height to contact length. Rectangular, isosceles and right-triangular elements of equal contact
length and height are compared from 53 to 67 GHz in full-wave openEMS simulations including the
waveguide-to-rod transitions. The rectangle peaks sharply at the rule, reaching 15.9 dBi. The right triangle
keeps its gain within 0.7 dB and its side lobes at least 10 dB down over a two-to-one height range. At the
rule, both outperform the isosceles element by 1.2 to 2.1 dB, and all remain matched.
\end{abstract}

\vskip0.5\baselineskip

\section{Introduction}
Pinching antennas (PAs) radiate from a low-loss dielectric waveguide by bringing a small dielectric
element into contact with it, so that part of the guided power is coupled out where the element is
placed~\cite{fukuda2022pinching}. System-level studies exploit the freedom to position and activate such
elements along the waveguide, and mostly model each PA as an isotropic point
source~\cite{ding2025flexible,wang2025modeling,tyrovolas2026ergodic,tyrovolas2026howmany}; directional
element patterns have only recently been included in the channel model~\cite{zhang2025directional}.
Li~\emph{et~al.}~\cite{li2026geometry} showed with full-wave simulations and measurements at 60\,GHz that
the pattern of a PA depends strongly on its shape: a regular triangle radiates a single, cleaner beam than a
square of the same thickness, which was attributed to reduced internal reflections and a tapered
polarization current. What is still missing is a design rule: which geometric feature of the element sets
its gain, beamwidth and beam direction, and how the element should be dimensioned.

This paper derives such a rule from a phase-matching picture of the PA and tests it with a systematic
full-wave comparison of three outlines that share the same contact length and height. The contributions are:
(i) we show that the PA, being denser than the guided mode, acts as a prism coupler~\cite{tien1970prism}:
power leaves the rod at an angle fixed by the effective index of the \emph{plate} mode of the element, not by
the bulk permittivity, and this angle predicts both the optimum element height and the beam direction;
(ii) we quantify how the contact length $L$ and the height $h$ of rectangular, isosceles and right-triangular
elements set the S-parameters, gain, beamwidth and beam direction;
(iii) we show that elements with a vertical downstream face sized by this rule outperform the isosceles
shape, and that a right-triangle outline makes the element tolerant to its height while keeping the side
lobes low; and (iv) we provide a model of the complete system in the open-source FDTD solver
openEMS~\cite{openems}, including the
circular-waveguide (CW) to rod transitions, which reproduces the main-lobe directions reported
in~\cite{li2026geometry}.

\section{System Model and Numerical Method}
\begin{figure}[!t]
\centering
\subfloat[]{\includegraphics[width=\columnwidth]{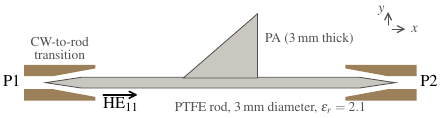}\label{fig:geom_sys}}\\
\subfloat[Rectangle]{\includegraphics[width=0.31\columnwidth]{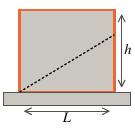}\label{fig:geom_rect}}
\hfil
\subfloat[Right triangle]{\includegraphics[width=0.31\columnwidth]{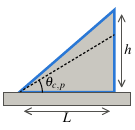}\label{fig:geom_rtri}}
\hfil
\subfloat[Isosceles]{\includegraphics[width=0.31\columnwidth]{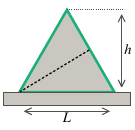}\label{fig:geom_iso}}
\caption{(a)~Simulated system (section in the PA plane, $z=0$): PTFE rod between two CW-to-rod transitions;
P1 is excited, P2 is matched. (b)--(d)~PA outlines compared at equal contact length $L$ and height $h$; the
dashed line is the in-plate phase-matched ray at $\theta_{c,p}$ launched from the upstream end of the contact.}
\label{fig:geom}
\end{figure}

The system follows~\cite{li2026geometry} (Fig.~\ref{fig:geom_sys}). A PTFE rod ($\varepsilon_r=2.1$,
$\tan\delta=2\times10^{-4}$, diameter \SI{3}{mm}) runs along $x$ and carries the HE$_{11}$ mode, the only
guided mode below \SI{72.9}{GHz}. Each end of the rod is fed by a copper transition in the spirit
of~\cite{vu2021bending}: an empty CW of \SI{3.6}{mm} diameter (TE$_{11}$ cut-off \SI{48.8}{GHz}) flares in a
\SI{12}{mm} conical horn to an \SI{8}{mm} aperture, and the rod enters it with a \SI{10}{mm} conical taper
whose tip lies \SI{2}{mm} inside the straight section. The transition is axisymmetric, so TE$_{11}$ couples
only to HE$_{11}$. Wave ports with analytic TE$_{11}$ profiles are placed in the straight sections; port~2
also records the cross-polarized TE$_{11}$ mode. The free rod between the horn apertures is \SI{70}{mm}
long, and the PA is centred on it.

The PA is a \SI{3}{mm}-thick PTFE plate standing on the rod in the $xy$ plane, with its base sunk
\SI{0.2}{mm} into the rod to model direct contact (a \SI{1.5}{mm}-wide contact strip). Three outlines are
compared at the same base (contact) length $L$ and height $h$ (Fig.~\ref{fig:geom_rect}--\ref{fig:geom_iso}):
a rectangle, an isosceles triangle with its apex above the middle of the base, and a right triangle whose
vertical leg is at the downstream end. The regular triangle of~\cite{li2026geometry} is the isosceles case
with $L=\SI{20.78}{mm}$, $h=\SI{18}{mm}$. The input TE$_{11}$ mode has its electric field along $y$, i.e.\ in
the plane of the plate.

The structure is solved with the FDTD solver openEMS~\cite{openems} over \SIrange{52}{70}{GHz}, with cells of
\SI{0.25}{mm} along the rod, \SI{0.15}{mm} across the rod and plate, and 8-cell PML boundaries (5.5--9~M cells
per model). The far field is computed from a Huygens box from which the two CW bores are removed, so that
guided power is not counted as radiation; gain is referred to the power accepted at P1. Radiation results are
reported over \SIrange{53}{67}{GHz}, where the unaccounted power $P_\mathrm{acc}-P_\mathrm{P2}-P_\mathrm{rad}$,
which includes the dielectric loss, stays within \SI{\pm 3.1}{\%} of $P_\mathrm{acc}$ for every element in
Table~\ref{tab:cmp} (and below \SI{6.5}{\%} for all models). Without a PA, the two transitions in series give
$|S_{11}|<\SI{-15.9}{dB}$ and $|S_{21}|$ between \SI{-0.43}{dB} and \SI{-0.17}{dB} over this band.

Angles are given as the beam direction $\psi$ in the PA plane ($\theta=\SI{90}{\degree}$), measured from the
rod axis toward the PA; the azimuth of~\cite{li2026geometry} is $\phi=\SI{270}{\degree}-\psi$. For the square
and regular-triangle PAs of~\cite{li2026geometry}, the model gives main lobes at $\phi=\SI{245}{\degree}$ and
\SI{248}{\degree} against \SI{252}{\degree} and \SI{240}{\degree} reported there, the secondary lobes of the
square in the same sector (\SIrange{98}{134}{\degree} here, \SIrange{108}{130}{\degree}
in~\cite{li2026geometry}), and the triangle again more directive than the square. Our absolute gains are
\SIrange{1.5}{3.2}{dB} lower; the transition and rod lengths of~\cite{li2026geometry} are not given, so this
offset is not interpreted further.

\section{Coupling Mechanism and Sizing Rule}
The HE$_{11}$ field outside the rod decays over about \SI{1.4}{mm} at 60\,GHz, and the mode is slow
($n_g=\beta_g/k_0=1.150$, with $\beta_g$ its propagation constant and $k_0$ the free-space wavenumber). A PTFE
element in contact is denser than this mode, so the guided wave cannot stay bound under it: as in a prism
coupler~\cite{tien1970prism}, power tunnels into the element as a wave whose phase velocity along $x$ matches
that of the rod. Inside a plate of thickness $t$ this wave is the fundamental plate mode (TE$_0$ for a field
in the plate plane) of effective index $n_p$~\cite{okamoto2006fundamentals}, and it travels at
\begin{equation}
\cos\theta_{c,p} = n_g/n_p
\label{eq:cherenkov}
\end{equation}
to the rod axis. For $t=\SI{3}{mm}$, $n_p=1.343$ and $\theta_{c,p}=\SI{31.1}{\degree}$ at 60\,GHz, whereas the
bulk index $\sqrt{2.1}$ would give \SI{37.5}{\degree}. Over \SIrange{53}{67}{GHz}, $\theta_{c,p}$ decreases from
\SI{33.4}{\degree} to \SI{28.8}{\degree}, because $n_g$ rises faster than $n_p$.

Equation~\eqref{eq:cherenkov} has two consequences. First, power is collected along the whole contact, so the
radiated fraction grows with $L$. Second, the launched wave must reach a face through which it can leave. A
wave at $\theta_{c,p}$ meets a horizontal top face \SI{58.9}{\degree} off its normal, beyond the critical angle
$\arcsin(1/n_p)=\SI{48.1}{\degree}$, and is turned back toward the rod; it meets a vertical downstream face
\SI{31.1}{\degree} off its normal and can leave. The wave launched at the upstream end of the contact reaches
the top of that face when
\begin{equation}
h = L\tan\theta_{c,p},
\label{eq:rule}
\end{equation}
which sets the element height for a given contact length.

\begin{figure}[!t]
\centering
\subfloat[Right triangle]{\includegraphics[width=\columnwidth]{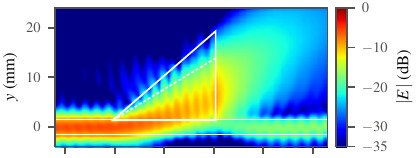}\label{fig:fields_rtri}}\\
\subfloat[Rectangle]{\includegraphics[width=\columnwidth]{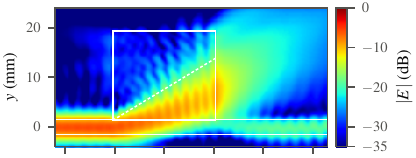}\label{fig:fields_rect}}\\
\subfloat[Isosceles]{\includegraphics[width=\columnwidth]{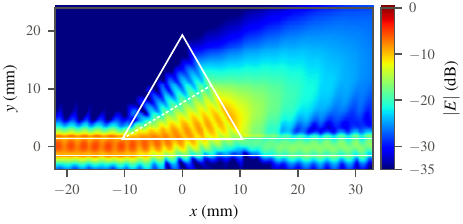}\label{fig:fields_iso}}
\caption{$|E|$ at 60\,GHz in the PA plane for $L=\SI{20.78}{mm}$, $h=\SI{18}{mm}$ (the size of the regular
triangle of~\cite{li2026geometry}), normalized to its maximum. Dashed: in-plate ray at $\theta_{c,p}$ from
the upstream end of the contact.}
\label{fig:fields}
\end{figure}

Fig.~\ref{fig:fields} shows this behaviour for the three outlines at the size of the regular triangle
of~\cite{li2026geometry}. In all cases the field leaves the rod along the contact and climbs into the plate
at an angle close to $\theta_{c,p}$. In the right triangle (Fig.~\ref{fig:fields_rtri}) the wave fills the plate
under the ramp and exits through the vertical leg as a single beam. In the rectangle (Fig.~\ref{fig:fields_rect}),
the region above the ray shows a standing-wave pattern: part of the wave is reflected inside the plate, which
broadens the beam. In the isosceles element (Fig.~\ref{fig:fields_iso}) the wave meets the slanted downstream
face close to the rod and leaves there, which tilts the beam toward the rod.

\section{Parametric Study}
\begin{figure*}[!tp]
\centering
\includegraphics{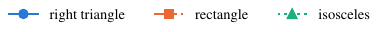}\\[-6pt]
\subfloat[Gain]{\includegraphics[width=0.24\textwidth]{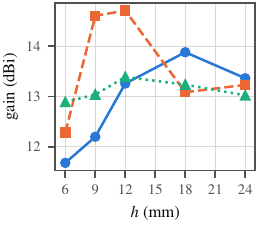}\label{fig:h_gain}}
\hfil
\subfloat[Beam direction]{\includegraphics[width=0.24\textwidth]{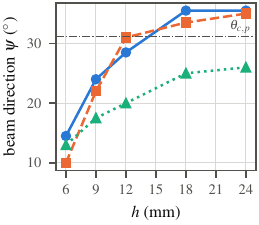}\label{fig:h_psi}}
\hfil
\subfloat[Half-power beamwidth]{\includegraphics[width=0.24\textwidth]{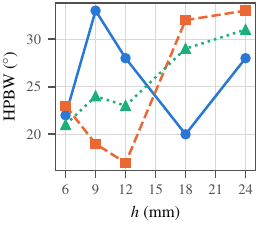}\label{fig:h_hpbw}}
\hfil
\subfloat[$|S_{11}|$ and $|S_{21}|$]{\includegraphics[width=0.24\textwidth]{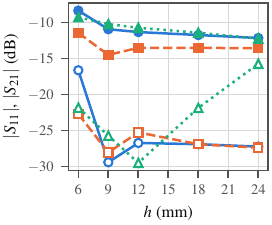}\label{fig:h_s}}\\
\subfloat[Gain]{\includegraphics[width=0.24\textwidth]{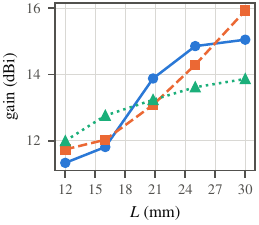}\label{fig:L_gain}}
\hfil
\subfloat[Beam direction]{\includegraphics[width=0.24\textwidth]{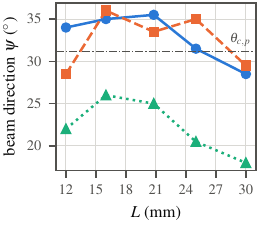}\label{fig:L_psi}}
\hfil
\subfloat[Half-power beamwidth]{\includegraphics[width=0.24\textwidth]{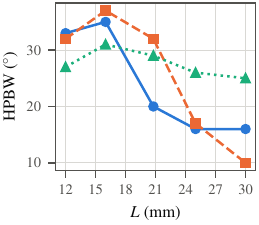}\label{fig:L_hpbw}}
\hfil
\subfloat[$|S_{11}|$ and $|S_{21}|$]{\includegraphics[width=0.24\textwidth]{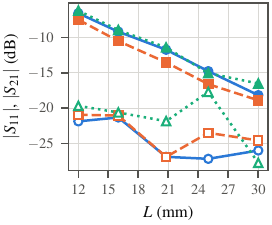}\label{fig:L_s}}
\caption{Effect of the height $h$ at $L=\SI{20.78}{mm}$ (a)--(d) and of the contact length $L$ at
$h=\SI{18}{mm}$ (e)--(h), all at 60\,GHz. In (d) and (h), filled markers show $|S_{21}|$ and open markers
$|S_{11}|$.}
\label{fig:sweeps}
\end{figure*}

Fig.~\ref{fig:sweeps} reports the effect of $h$ and $L$ on the three outlines at 60\,GHz. All 33 simulated
elements have $|S_{11}|$ between \SI{-32.4}{dB} and \SI{-15.7}{dB} at 60\,GHz and stay below \SI{-10}{dB}
over the whole simulated band, so their \SI{-10}{dB} fractional bandwidth exceeds \SI{29.5}{\%}; it is limited
by the simulated band, not by the element. $|S_{21}|$ falls with $L$ as more power is radiated: short elements
($L\leq\SI{16}{mm}$) radiate \SIrange{75}{91}{\%} of the accepted power, and all elements radiate
\SI{98}{\%} at $L=\SI{30}{mm}$.

At fixed $L=\SI{20.78}{mm}$ (Fig.~\ref{fig:h_gain}--\ref{fig:h_s}), the rectangle is best for
$h=\SIrange{9}{12}{mm}$ (\SIrange{14.6}{14.7}{dBi}, \SIrange{17}{19}{\degree} beamwidth) and degrades when made
taller; this optimum brackets $L\tan\theta_{c,p}=\SI{12.5}{mm}$ from~\eqref{eq:rule}. The right triangle reaches
its maximum for $h=\SIrange{12}{18}{mm}$, i.e.\ for ramp angles from about $\theta_{c,p}$ to
$\theta_{c,p}+\SI{10}{\degree}$, while the isosceles element stays within \SIrange{13.0}{13.4}{dBi} for
$h\geq\SI{9}{mm}$. For the two elements with a vertical downstream face the beam rises from about
\SI{10}{\degree} to \SI{35}{\degree} as $h$ grows and saturates at \SIrange{33.5}{35.5}{\degree}, a few degrees
above $\theta_{c,p}$, whereas the isosceles beam stays \SIrange{13}{26}{\degree} from the axis.

At fixed $h=\SI{18}{mm}$ (Fig.~\ref{fig:L_gain}--\ref{fig:L_s}), short elements ($h>L\tan\theta_{c,p}$) are
poor for every outline. As $L$ grows, the gain of the two elements with a vertical face keeps rising, to
\SI{15.1}{dBi} (right triangle) and \SI{15.9}{dBi} (rectangle) at $L=\SI{30}{mm}$, where~\eqref{eq:rule} is
met, while the isosceles element saturates at \SI{13.9}{dBi}.

\begin{figure*}[!tp]
\centering
\includegraphics{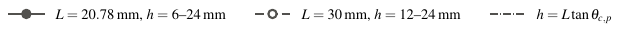}\\[-6pt]
\subfloat[Rectangle]{\includegraphics[width=0.24\textwidth]{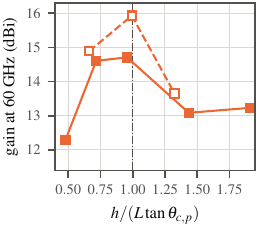}\label{fig:rule_rect}}
\hfil
\subfloat[Right triangle]{\includegraphics[width=0.24\textwidth]{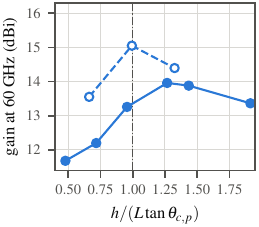}\label{fig:rule_rtri}}
\hfil
\subfloat[Isosceles]{\includegraphics[width=0.24\textwidth]{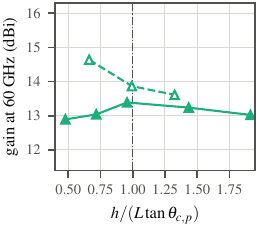}\label{fig:rule_iso}}
\hfil
\subfloat[Beam direction at the rule]{\includegraphics[width=0.24\textwidth]{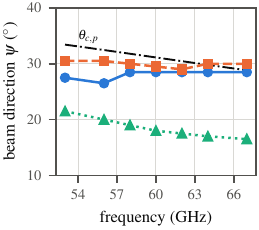}\label{fig:rule_dir}}\\
\subfloat[Gain pattern at the rule, 60\,GHz]{\includegraphics[width=0.42\textwidth]{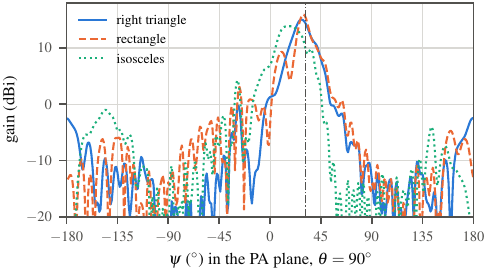}\label{fig:patterns}}
\hfil
\begin{minipage}[b]{0.54\textwidth}
\centering
\subfloat[Right triangle]{\includegraphics[width=0.31\linewidth]{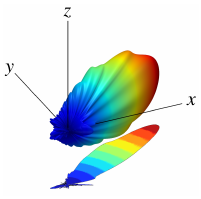}\label{fig:3d_rtri}}
\hfil
\subfloat[Rectangle]{\includegraphics[width=0.31\linewidth]{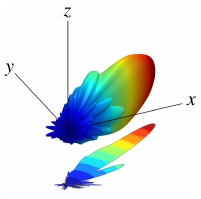}\label{fig:3d_rect}}
\hfil
\subfloat[Isosceles]{\includegraphics[width=0.31\linewidth]{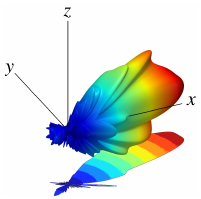}\label{fig:3d_iso}}
\\[2pt]
\includegraphics[width=0.75\linewidth]{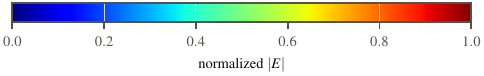}\\%
\end{minipage}
\caption{(a)--(c)~Gain at 60\,GHz versus the normalized height $h/(L\tan\theta_{c,p})$ for the two height
sweeps at fixed contact length; the dash-dotted line marks~\eqref{eq:rule}. (d)~Beam direction versus
frequency, (e)~gain in the PA plane ($\theta=\SI{90}{\degree}$) at 60\,GHz, and (f)--(h)~3D normalized
$|E|$ at 60\,GHz, all for $L=\SI{30}{mm}$, $h=\SI{18}{mm}$; the dash-dotted lines mark~\eqref{eq:cherenkov}, and
in (f)--(h) $x$ is the rod axis (propagation direction), the PA lies in the $xy$ plane toward $+y$, and the
pattern in that plane is projected below each beam as filled contours.
Colours and markers as in Fig.~\ref{fig:sweeps}.}
\label{fig:rule}
\end{figure*}

Fig.~\ref{fig:rule} tests~\eqref{eq:rule} with the two height sweeps at fixed contact length,
$L=\SI{20.78}{mm}$ and $L=\SI{30}{mm}$, plotted against the normalized height $h/(L\tan\theta_{c,p})$. For the
rectangle (Fig.~\ref{fig:rule_rect}) both sweeps peak at a normalized height of 1: \SI{14.7}{dBi} at 0.96 for
$L=\SI{20.78}{mm}$ and \SI{15.9}{dBi} at 1.00 for $L=\SI{30}{mm}$, with \SIrange{1.6}{2.3}{dB} less at
\SIrange{1.33}{1.44}{}. The right triangle (Fig.~\ref{fig:rule_rtri}) also needs $h\gtrsim L\tan\theta_{c,p}$,
losing \SI{1.8}{dB} at 0.72 ($L=\SI{20.78}{mm}$) and \SI{1.5}{dB} at 0.66 ($L=\SI{30}{mm}$), but above the
rule it is tolerant: its gain varies by only \SI{0.7}{dB} from 0.96 to 1.92 at $L=\SI{20.78}{mm}$ and drops
by \SI{0.7}{dB} at 1.33 for $L=\SI{30}{mm}$, against \SI{2.3}{dB} for the rectangle, and its side lobes stay
at or below \SI{-10}{dB} at all these heights. The ramp removes the plate region above the phase-matched wave
in which the rectangle traps reflected power. The isosceles element (Fig.~\ref{fig:rule_iso}) has no maximum at
the rule: its gain stays within \SIrange{12.9}{13.4}{dBi} at $L=\SI{20.78}{mm}$ and decreases with height at
$L=\SI{30}{mm}$, from \SI{14.6}{dBi} at 0.66 to \SI{13.6}{dBi} at 1.33, while its beam stays
\SIrange{13}{26}{\degree} from the axis.

Fig.~\ref{fig:rule_dir} shows that, at the rule, the beam of the rectangle and of the right triangle stays
within \SI{3}{\degree} and \SI{6}{\degree} of $\theta_{c,p}$ over \SIrange{53}{67}{GHz}, whereas the isosceles
element radiates \SIrange{12}{13}{\degree} closer to the rod. The phase-matching angle therefore sets both the
size of a good element and the direction of its beam.

\begin{table}[!t]
\centering
\caption{Comparison at Equal $L$ and $h$ at 60\,GHz}
\label{tab:cmp}
\footnotesize
\setlength{\tabcolsep}{2.6pt}
\begin{tabular}{@{}lccccccc@{}}
\toprule
 & $G$ & HPBW & SLL & $\psi$ & $\eta$ & $|S_{11}|$ & $|S_{21}|$\\
 & (dBi) & (\si{\degree}) & (dB) & (\si{\degree}) & (\%) & (dB) & (dB)\\
\midrule
\multicolumn{8}{@{}l}{$L=\SI{20.78}{mm}$, $h=\SI{18}{mm}$}\\
Right tri. & 13.9 & 20 & -10.2 & 35.5 & 93 & -26.9 & -11.8\\
Rectangle & 13.1 & 32 & -10.0 & 33.5 & 95 & -26.9 & -13.5\\
Isosceles & 13.2 & 29 & -8.2 & 25.0 & 93 & -21.9 & -11.4\\
\midrule
\multicolumn{8}{@{}l}{$L=\SI{30}{mm}$, $h=\SI{18}{mm}$}\\
Right tri. & 15.1 & 16 & -10.7 & 28.5 & 98 & -26.0 & -18.2\\
Rectangle & 15.9 & 10 & -6.6 & 29.5 & 98 & -24.6 & -18.9\\
Isosceles & 13.9 & 25 & -9.6 & 18.0 & 98 & -27.7 & -16.5\\
\bottomrule
\end{tabular}\\[2pt]
\parbox{\columnwidth}{\scriptsize $\eta$: radiated/accepted power.}
\end{table}

Table~\ref{tab:cmp} and Fig.~\ref{fig:patterns}--\ref{fig:3d_iso} compare the three outlines at the size of~\cite{li2026geometry}
and at $L=\SI{30}{mm}$, $h=\SI{18}{mm}$, which satisfies~\eqref{eq:rule}. At the size of~\cite{li2026geometry}
($h/(L\tan\theta_{c,p})=1.43$, above the rule) the right triangle already has the highest gain
(\SI{13.9}{dBi} against \SIrange{13.1}{13.2}{dBi}) and the narrowest beam (\SI{20}{\degree} against
\SIrange{29}{32}{\degree}), and, like the rectangle, a lower $|S_{11}|$ than the isosceles element
(\SI{-26.9}{dB} against \SI{-21.9}{dB}). At the rule, the rectangle reaches \SI{15.9}{dBi} and a
\SI{10}{\degree} beam, but with \SI{-6.6}{dB} side lobes; the right triangle gives \SI{15.1}{dBi}, a
\SI{16}{\degree} beam and \SI{-10.7}{dB} side lobes with half the plate area. Both exceed the isosceles element
(\SI{13.9}{dBi}, \SI{25}{\degree}) by \SIrange{1.2}{2.1}{dB}, and point their beam \SIrange{10.5}{11.5}{\degree}
further from the rod. The isosceles element does better when it is lower (\SI{14.6}{dBi} at $h=\SI{12}{mm}$
for $L=\SI{30}{mm}$), but the right triangle and the rectangle at the rule still exceed it by \SI{0.4}{dB} and
\SI{1.3}{dB}.

\section{Conclusion}

A pinching-antenna element on a dielectric rod behaves as a prism coupler: the guided wave enters the denser
plate at the phase-matching angle $\theta_{c,p}$ of the plate mode, \SI{31}{\degree} for a \SI{3}{mm} PTFE
plate at 60\,GHz, and radiates through the face it reaches first. For elements with a vertical downstream face
this gives the sizing rule $h\approx L\tan\theta_{c,p}$, which locates the gain maximum of the rectangular element
and the onset of the gain plateau of the right-triangular one at both simulated contact lengths, while the beam points within \SI{6}{\degree} of $\theta_{c,p}$ across \SIrange{53}{67}{GHz}.
A right-triangle element, whose ramp removes the plate region above the phase-matched wave, trades
\SI{0.9}{dB} of gain against the rectangle for tolerance to its height and side lobes at or below
\SI{-10}{dB}, and both shapes outperform the isosceles element of equal size by \SIrange{1.2}{2.1}{dB}. These
results give a simple, physically grounded starting point for designing PAs with a prescribed gain and beam
direction, and the fabrication and measurement of the proposed elements will follow.

\section*{Acknowledgment}
Generative AI tools were used solely for language editing, grammar
correction, and improving the clarity of the text. All scientific content,
methodology, analysis, results, and conclusions were developed, verified, and
approved by the authors.

\bibliographystyle{IEEEtran}
\bibliography{refs}

\end{document}